\documentclass{elsart}

\usepackage{graphicx,amssymb}
\journal{Physica D}

\newcommand{\I}{\mathrm{i}}
\newcommand{\D}{\mathrm{d}}

\begin{document}
\begin{frontmatter}
\title{On the Doppler distortion of the sea-wave spectra.}
\author[LandauITP]{A.O. Korotkevich\corauthref{cor}},
\corauth[cor]{Corresponding author.}
\ead{kao@itp.ac.ru}
\address[LandauITP]{Landau Institute for Theoretical Physics,
Kosygin Str. 2, Moscow, 119334, Russian Federation}

\begin{abstract}

Discussions on a form of a frequency spectrum of wind-driven sea waves
just above the spectral 
maximum have continued for the last three decades. In 1958 Phillips made a
conjecture that wave 
breaking is the main mechanism responsible for the spectrum formation \cite{Phillips1958}. 
That leads to the spectrum decay $\sim \omega^{-5}$, where $\omega$ is the frequency of waves.
There is a contradiction
between the numerous experimental data and this 
spectrum. Experiments frequently show decay $\sim\omega^{-4}$ \cite{Toba1973},\cite{Donelan1985},\cite{Hwang1999}. There are several
ways of the explanation of this phenomenon. One of them
(proposed by Banner \cite{Banner1990}) takes into account the Doppler effect due to surface
circular currents generated by underlying waves in the Phillips model.

In this article the influence of the Doppler effect on an arbitrary averaged spectrum is considered
using both analytic and numerical approaches. Although we mostly concentrated on the very important
case of Phillips model the developed technique and general formula can be used for the analysis of other spectra.

For the particular case of Phillips spectra we got analytic asymptotics in the vicinity of spectral maximum
and for high frequencies. Results were obtained for two most important angular dependences of the spectra:
isotropic and strongly anisotropic. Together with the analytic investigation we performed numerical calculations
in a wide range of frequencies. Both high and low frequency asymptotics are in very good agreement with
the numerical results.

It was shown that at least at low frequencies correction to the spectrum due to the Doppler shift is negligible.
At high frequencies there is an asymptotic with tail $\sim \omega^{-3}$.
\end{abstract}

\end{frontmatter}

\section{Introduction}
There is a long discussion in the scientific world about the form of wind generated sea wave spectrum in the 
"universal" region of frequencies $\omega \gtrsim \omega_p$, where $\omega_p$ is the frequency of the 
spectral maximum.

There are at least two main points of view. The first one proposed by O.\,Phillips in 1958 \cite{Phillips1958} gives us the Phillips' spectrum in 
universal region
\begin{equation}
\label{Phil-1}
F(\omega) \simeq \frac{\alpha g^2}{\omega^5},
\end{equation}
where $\alpha$ --- is a dimensionless constant, which according to experimental data appears to be rather
small $\alpha \simeq 0.01$.
The second point of view introduced by V.E. Zakharov and N.N. Filonenko \cite{Zakharov1966} gives us in universal region 
following spectrum
\begin{equation}
\label{ZakhFil}
F(\omega) \simeq \frac{\beta \varepsilon^{\frac{1}{2}} g v}{\omega^4},
\end{equation}
where $\beta$ --- is a dimensionless constant, $v$ --- is wind velocity, $\varepsilon \simeq \rho_a / \rho_w$ - 
is the ratio of ocean water and atmosphere densities.

The formulae (\ref{Phil-1}) and (\ref{ZakhFil}) are based upon completely opposite propositions.
The Phillips' spectrum (\ref{Phil-1}) takes place if the spectrum at the high frequency region is mainly
determined by appearance of sharp crests and wave breaking, which are strongly nonlinear phenomena.

On the other side spectrum (\ref{ZakhFil}) is based on presumption of small mean steepness ---
 $\mu \ll 1$ and based on weakly nonlinear waves interaction. Usually, for mature sea $\mu \simeq 0.1$.
 
 Following this ideology the wind generated waves' ensemble is described by Hasselmann kinetic equation 
 \cite{Hasselmann1962}, \cite{Hasselmann1963}
 \begin{equation}
 \label{Hasselmann}
 \frac{\partial N}{\partial t} = S_{nl} + p^{+} + p^{-}. 
 \end{equation}
 Zakharov-Filonenko spectrum appears as an exact solution of the particular case of Eq. (\ref{Hasselmann})
 \begin{equation}
 S_{nl} = 0,
 \end{equation}
this is a classical case of Kolmogorov-Zakharov (KZ) spectra. The theory of weak turbulence is far 
advanced both analytically and numerically \cite{Zakharov1966}-\cite{ZKPR2007}.

In the region of moderate frequencies in field experiments with mature sea
the $\omega^{-4}$ spectrum is dominating with confidence \cite{Toba1973}-\cite{Hwang1999}.
From this fact the question of physical interpretation of spectrum $F(\omega) \simeq \omega^{-4}$
becomes very important. Following Zakharov and Filonenko this
spectrum is just weakly turbulent KZ-spectrum. There are other explanations, though.

One of them was proposed by Phillips in 1986 \cite{Phillips1986}. He made the supposition that spectrum $\omega^{-4}$ is
a result of a balance of all three terms of Eq.(\ref{Hasselmann})
\begin{equation}
\label{Phil-2}
S_{nl} \simeq p^{+} + p^{-}.
\end{equation}
This hypothesis is unlikely to be treated as a theory, because the expression of $p^{-}$ is almost
completely unknown. There are only very rough empirical formulas. However in that paper this
expression was taken in the form that gives us the balance in Eq. (\ref{Phil-2}).
For now there are no strong argument in favor of this conjecture.

Another explanation was proposed by Banner \cite{Banner1990} and developed by Donelan \cite{Donelan1998}. 
They considered the $\omega^{-4}$ spectrum as "an artifact of the 
observation of time histories at a point brought about by Doppler shifting of the short waves 
riding on the orbital currents of the long waves" --- citation from \cite{Donelan1998}.

In the paper \cite{Banner1990} yet another explanation was proposed. It is based on taking
into account the angle dependence of spatial spectra. Today there is no common opinion about the exact
form of such a distribution. For example
angular distributions obtained by Donelan {\it et al} \cite{Donelan1985} and Hwang {\it et al} \cite{Hwang1999}
in field experiments differ significantly. Also, the universality of angular dependence is not obvious yet.

The main goal of this paper is to find a frequency spectra of waves from known space-spectra in the presence
of the Doppler effect.
Let us suppose that spatial spectrum of the short waves $\Phi(k)$ is known. The problem is to derive the 
frequency spectrum $F(\omega)$ using only $\Phi(k)$. Normally we have to assume that waves obey the same
dispersion relation as in the linear case. Thus the following simple expression takes place for
positive-frequency part of the energy spectrum
$$
F(\omega) = \int \Phi(k) \delta (\omega-\omega(k)) \D k.
$$
However in the presence of long waves the dispersion relation of the short waves modifies due to Doppler
effect at the orbital velocity's field ${\vec v}$ of the long wave component
\begin{equation}
\omega(k) \rightarrow \omega(k) + ({\vec k} {\vec v}).
\end{equation}
In the well-known article by Kitaigorodskii, Krasitskii, Zaslavskii \cite{KKZ1975} it was considered that ${\vec v}$
is a random function on time. In fact, it is a quasi-periodic function with random envelope. In the present
paper we simplify the problem and assume that ${\vec v}$ is a pure periodic function. Another word, we
study modification of the spectrum due to a presence of a long monochromatic wave. This assumption gives
us an opportunity to develop the analytical theory up to explicit formulae.

The present paper is devoted to the quantitative theory of this without any doubt very important
phenomenon. We consider it in a very simplified way, supposing that ${\vec v}$ is a sinusoidal function
of time, i.e. assuming that the long wave is strictly monochromatic. In spite of simplicity
this assumption no doubt gives us the correct order of magnitude.

In Sect. 2
we introduced the common notations (following Phillips). In Sect. 3 the
general formula for spectrum
(taking into account Doppler shifting) is derived. It is
convenient to introduce a new set of dimensionless variables,
this is considered in Sect. 4.
In Sect. 5 we develop an important case of dispersion $\omega^2 = g k$ ("deep
water" surface gravity waves) where
g is the gravity acceleration, k is the wave number equal to $2 \pi/\lambda$.
Here we also apply developed technique to the important case of Phillips model.
Section 6  is devoted to
numerical results. The conclusion is placed in Sect. 7. 

In the case of relatively small background velocities (with respect to phase velocity) the
Doppler effect appears to be too weak (the first correction to the Phillips spectrum
is proportional to $\sim (v \omega/g)^2$) in a wide range of frequencies. In an opposite case of high speed
background Doppler shifting is very important and gives us at high frequencies another asymptotic tail $\sim \omega^{-3}$. Numerical results are in good agreement with analytic asymptotics and give us
information about spectra distortion in the intermediate range of frequencies.

\section{Formulation of the problem and notations}
Following the notations of Phillips, the wave spectrum for homogeneous
stationary wave field can be 
introduced as follows
\begin{equation}
\label{spectrum}
X(\vec k, \omega) = (2 \pi)^{-3} 
\int\!\!\!\int\limits_{-\infty}^{+\infty}\rho(\vec r, t) \exp [-i(\vec
k \vec r - \omega t)] \D\vec r \D t,
\end{equation}
 where $\rho (\vec r, t) = \overline{\xi (\vec x,t_0)\xi(\vec x +
\vec r , t_0 +t)}$ is the 
 covariance of the surface displacement $\xi(\vec x,t)$, $\vec r$ is
the spatial separation vector, $t$
 is the time separation, $\vec k = \overrightarrow
 {(k_1,k_2)} =  \overrightarrow {(k, \theta)}$ is the wavenumber vector (the 
 second expression is the wavevector respresentation in polar coordinates) and $\omega$ is
the radian frequency.
 Obviously it can be rewritten in terms of Fourier transforms of corresponding
functions
\begin{equation}
\left< \xi (\vec k,\omega) \xi^{*} (\vec {k'},\omega')\right> = 
X (\vec k,\omega) \delta_{\vec k- \vec {k'}}
\delta_{\omega-\omega'},
\end{equation}
$$
X(-\vec k,-\omega) = X(\vec k,\omega).
$$
It is convenient to introduce additional spectra which are reduced forms
of Eq. (\ref{spectrum})
\begin{equation}
\Phi(\vec k) = 2 \int\limits_{0}^{+\infty} X(\vec k,\omega) \D \omega, 
\end{equation}
\begin{equation}
\label{F_def}
F(\omega) = 2 \int\limits_{0}^{2\pi}\int\limits_{0}^{+\infty} X(\vec
k,\omega) k \D k \D\theta.
\end{equation}
In general case it is impossible to express $F(\omega)$ in terms of $\Phi
(\vec k)$.

There are many complex mechanisms of long and short waves interaction.
They were considered by Phillips \cite{Phillips1981} at the vicinity of spectral peak frequency $\omega_p$.
According to this paper, the first correction terms taking into account variable effective gravity and other nonlinear
effects are proportional to steepness (which is supposed to be small in our consideration).
In our case, when the Doppler velocity does not depend on steepness, the Doppler effect is also insensitive to
roughness of the waves. In the special case, when we consider influence of a periodic current due to
spectral peak we discuss this issue in details.
In this article we limit our consideration only by Doppler effect. In this sense our model is different with
respect to one used by Banner \cite{Banner1990}. Nevertheless, the results obtained are in an agreement
with experimental data by Donelan \cite{Donelan1985} which show no or indistinguishable dependence of spectral
slope with moderate change of steepness. The more detailed consideration can be an inspiration for future
investigations.

Following Appendix~\ref{appendix_0} we shall use linearized relations for spectra. For an isotropic spectrum
we have~(\ref{Chi_Phi_iso})
\begin{equation}
X_{is} (\vec k,\omega) = \frac{1}{2}\Phi (k) \left( \delta (\omega - \omega_k) +
\delta (\omega + \omega_k) \right),
\end{equation}
where $\omega_k = \omega (k)$ is the dispersion law and $\delta$ is the Dirac
delta-function.

In the weakly nonlinear approximation we consider interaction between positive and negative-frequency
parts of the spectra to be negligible. This simplification let us to find dependence
between spatial and frequency spectra. From now on we shall consider positive-frequency parts of spectra.

In this case in the isotropic case we shall get
\begin{equation}
X_{is} (\vec k,\omega) = \frac{1}{2}\Phi (k) \delta (\omega - \omega_k).
\end{equation}

In a unidirectional (or strongly anisotropic) case, when all waves propagate in one direction,
corresponding to angle $\theta_0$, the relation is a little bit different~(\ref{Chi_Phi_aniso})
\begin{equation}
X_{an} (\vec k,\omega) = \Phi (k) \delta(\theta-\theta_0) \delta (\omega - \omega_k).
\end{equation}

From now on we shall, for the sake of
universality, hide the multiplier $1/2$ in isotropic case in the angular dependence of the spectrum.
Finally, we get simplified relation for the spectra
\begin{equation}
X (\vec k,\omega) = \Phi (\vec k) \delta (\omega - \omega_k).
\end{equation}

Now one can rewrite Eq. (\ref{F_def}) in the following form
\begin{equation}
\label{F_def_linear}
F(\omega) = 2 \int\limits_{0}^{+\infty} \int\limits_{0}^{2 \pi} \Phi(k,\theta)
\delta (\omega - \omega_k) \D\theta k \D k.
\end{equation}
Using the rules of integration of $\delta$-function one can obtain
\begin{equation}
\begin{array}{l}
\displaystyle
F(\omega) = \frac{2k}{|\omega'(k)|} \int\limits_{0}^{2 \pi} \Phi(k, \theta) \D
\theta,\\
\displaystyle
k = k (\omega),
\end{array}
\end{equation}
where $k(\omega)$ is the inverse function for $\omega(k)$.

In the deep water case $\omega = \sqrt {gk}$
$$
k=\frac{\omega^2}{g}, \omega'(k) = \frac{1}{2} \sqrt{\frac{g}{k}}=\frac{1}{2}
\frac{g}{\omega}
$$
\begin{equation}
F(\omega) =\frac{4 \omega^3}{g^2}\int\limits_{0}^{2 \pi}
\Phi\left(\frac{\omega^2}{g},\theta \right) \D\theta.
\end{equation}
For an isotropic spectrum ($\Phi(k,\theta) = \Phi(k)/2$)
\begin{equation}
\label{i_spectrum_begin}
F_{is}(\omega) = \frac{4 \pi \omega^3}{g^2}\Phi\left(\frac{\omega^2}{g} \right).
\end{equation}
For a strongly anisotropic spectrum ($\Phi(k,\theta) = \Phi(k) \delta(\theta
- \theta_0)$)
\begin{equation}
\label{sa_spectrum_begin}
F_{an}(\omega) = \frac{4 \omega^3}{g^2}\Phi\left(\frac{\omega^2}{g} \right).
\end{equation}
If the spectrum is determined by discontinuities of spatial derivative
(wedges)
caused by wave breaking according to Phillips \cite{Phillips1958}, we have
\begin{equation}
\label{Phillips_theory}
\xi \sim \frac{1}{k^2}, \Phi(k) \sim \frac{1}{k^4}.
\end{equation}
In the isotropic case substitution Eq. (\ref{Phillips_theory}) to Eq.
(\ref{i_spectrum_begin}) gives
\begin{equation}
\label{Phillips-isotropic}
F_{is0}(\omega) = 4 \pi \alpha \frac{g^2}{\omega^5}.
\end{equation}
Similar calculation for Eq. (\ref{sa_spectrum_begin}) gives us the following
result
\begin{equation}
\label{Phillips-anisotropic}
F_{an0}(\omega) = 4 \alpha \frac{g^2}{\omega^5}.
\end{equation}
Here $\alpha$ is the so-called Phillips' constant.

There is a recent work by Kuznetsov~\cite{Kuznetsov2004} which gives us
different spatial spectra for waves wedges due to more accurate analysis.
In this work we shall use spectrum~(\ref{Phillips_theory}) as an example,
because spectra of this type are still widely used in oceanology.
It should be stressed, that general formulae given below are valid for the Doppler distortion
of any waves spectra in a weakly nonlinear approximation.

\section{General formula for spectrum in the presence of Doppler effect}
Let us assume that we study waves on the background of periodic current
$\vec v (t)$ beneath 
the surface. Taking into account the Doppler effect in the dispersion
equation one can get the 
following substitution
$$
\omega(\vec k) \longrightarrow \omega (\vec k) + \vec {k v} (t).
$$
Suppose that this current is described by a one-dimensional periodic
function
$$
v(t) = v \cos(\omega_0 t).
$$
This allows us to rewrite our substitution in the following form
\begin{equation}
\label{Doppler_shift}
\omega(k) \longrightarrow \omega(k) + kv \cos (\omega_0 t) \cos \psi ,
\end{equation}
where $\psi$ is the angle formed by $\vec k$ and $\vec v$ vectors. 
Without loss of generality one can take $\psi = \theta$.

We can now find a range of velocities when we can consider only positive frequencies. In general
such conditions gives us $v \le w(k)/k = c_p$, where $c_p$ is the phase velocity. In the special case
of the Doppler shift due to spectral peak currents, using (\ref{Doppler_velocity_peak}) one can get
$$
k A_p \omega_p \le \omega.
$$
Which gives us
\begin{equation}
\omega \le \frac{1}{\mu}\omega_p, 
\end{equation}
where $\mu = k_p A_p$ is a steepness of spectrum peak wave. For the usual value $\mu=0.1$ we get
$\omega \le 10\omega_p$ which is more then enough for interpretation any open sea experimental data.
Even in the case of quite rough sea $\mu=0.2$ we cover frequency bandwidth of all current open water experiments.

Using expression (\ref{Doppler_shift}) one can write average value of expression (\ref{F_def_linear}) as
\begin{equation}
F(\omega) =\frac{2}{T} \int\limits_{0}^{T} \int\limits_{0}^{+\infty}
\int\limits_{0}^{2 \pi} \Phi (k,\theta) \delta (\omega - \omega(k) 
- kv \cos (\omega_0 t) \cos \theta) \D\theta k \D k \D t,
\end{equation}
where $T$ is the period equal to $2\pi/\omega_0$.
It is useful to take the average value of this expression with respect to $t$ as the first step. Let
us denote this factor to the part 
of the integral as $M$
$$
F(\omega) =\int\limits_{0}^{+\infty}
\int\limits_{0}^{2 \pi} \Phi (k,\theta) M (k, \theta) \D\theta k \D k,
$$
$$
M (k, \theta) =\frac{1}{\pi} \int \limits_{0}^{T} \delta (\omega - kv \cos \omega_0 t \cos
\theta - \omega_k)\D(\omega_0 t).
$$
The $\delta$-function in this expression gives us an equation for $t_0$
\begin{equation}
\omega - \omega_k = kv \cos \theta \cos (\omega_0 t_0),
\end{equation}
This equation have the following roots
\begin{equation}
\label{roots}
\omega_0 t_0 = \pm \arccos \frac{\omega - \omega_k}{kv \cos \theta},
\end{equation}
when the following inequality is satisfied
\begin{equation}
\label{inequality}
\left| \frac{\omega - \omega_k}{kv \cos \theta} \right| < 1.
\end{equation}
Using (\ref{roots}) and delta function integration rules one can get the
function M
\begin{equation}
M (k, \theta)=\frac{2}{\pi \left| kv \cos \theta \sqrt{1-\frac{(\omega - \omega_k)^2}{k^2 v^2
\cos^2 \theta}}\right|} =
\frac{2}{\pi \sqrt{k^2 v^2 \cos^2 \theta - (\omega - \omega_k)^2}}.
\end{equation}
Using this expression one can get
\begin{equation}
\label{spectrum_dim}
F(\omega) = \frac{2}{\pi} \int \int
\frac{\Phi (k,\theta) \D\theta k \D k}{\sqrt{k^2 v^2 \cos^2 \theta - (\omega -
\omega(k))^2}}.
\end{equation}
The limits of integration have to be chosen in order to satisfy inequality (\ref{inequality}).

\section{Convenient set of dimensionless variables}
For further consideration it is convenient to introduce a new variable and a parameter
\begin{equation}
\zeta = \frac{k v^2}{g}, \; \lambda = \frac{v \omega}{g}.
\end{equation}
One can see that $\zeta$ is a dimensionless analogue of $k$, and $\lambda$ is an analog of
$\omega$.
In these variables the spectrum Eq. (\ref{spectrum_dim}) takes the following
relatively simple form
\begin{equation}
\label{spectrum_dimless}
F(\lambda)=\frac{2}{\pi} \frac{g}{v^3} B \int\limits_{0}^{+\infty}
\int\limits_{0}^{2\pi} 
\frac{\Phi(\zeta, \theta) \D\zeta \D\theta}{\sqrt{\cos^2 \theta -
\left(\frac{\lambda - \lambda_\zeta}{\zeta} \right)^2}},
\end{equation}
where $B$ is the dimensional constant which has an origin in substitution $k \to \zeta$
in $\Phi(k)$. 
For instance in the Phillips spectrum case Eq. (\ref{Phillips_theory}) one can get
$$
\Phi(k) = \alpha k^{-4} = \alpha \left(\frac{\zeta g}{v^2}\right)^{-4} = \alpha \frac{v^8}{g^4} \zeta^{-4} = B \zeta^{-4},
$$
thus the constant takes the form $B=\alpha v^8 / g^4$.

The physical meaning of the variable $\lambda$ becomes more clear if we
transform it to the following form
\begin{equation}
\label{lambda_sence}
\lambda = \frac{\omega}{\omega_p} \frac{v}{c_p},
\end{equation}
where $\omega_p = g/c_p$ is the peak-frequency (this is the only
frequency-like scale in the model) 
and $c_p$ is the phase velocity corresponding to the spectral peak.

For the case of strongly anisotropic spectrum ($\Phi(k,\theta) =
\Phi(k)\delta(\theta)$) one can easily obtain 
the following result by a simple substitution in Eq. (\ref{spectrum_dimless})
\begin{equation}
\label{anisotr_spectrum_init}
F_{an}(\lambda)=\frac{2}{\pi} \frac{g}{v^3} B \int\limits_{0}^{+\infty}
\frac{\Phi(\zeta) \D\zeta}
{\sqrt{1-\left(\frac{\lambda - \lambda_\zeta}{\zeta} \right)^2}}.
\end{equation}

If the spectrum is isotropic ($\Phi(k,\theta) = \Phi(k)/2$) things become a
little bit more complex. Following 
Eq. (\ref{spectrum_dimless}) we have
\begin{equation}
\label{isotr_spectrum_init_1}
F_{is}(\lambda)=\frac{4}{\pi} \frac{g}{v^3} B \int\limits_{0}^{+\infty}
\Phi(\zeta) \D\zeta
\int\limits_{0}^{\pi/2} \frac{\D\theta}{\sqrt{\cos^2 \theta -
\left(\frac{\lambda - \lambda_\zeta}{\zeta} \right)^2}},
\end{equation}
An analysis of the $\theta$-part gives (Gradshteyn,Ryzhik)
$$
\int\limits_{0}^{\pi/2} \frac{\D\theta}{\sqrt{\cos^2 \theta -
\left(\frac{\lambda - \lambda_\zeta}{\zeta} \right)^2}}=
$$
$$
=\frac{1}{\sqrt{1-\left(\frac{\lambda - \lambda_\zeta}{\zeta} \right)^2}} 
K \left( \frac{1}{\sqrt{1-\left(\frac{\lambda - \lambda_\zeta}{\zeta} \right)^2}}
\right)=
$$
\begin{equation}
\label{isotr_K_part}
=K \left(\sqrt{1-\left(\frac{\lambda - \lambda_\zeta}{\zeta} \right)^2} \right) +
\frac{\I}{\sqrt{1-\left(\frac{\lambda - \lambda_\zeta}{\zeta} \right)^2}}
K\left(\frac{\lambda - \lambda_\zeta}{\zeta}\right),
\end{equation}
where $K$ is the complete elliptic integral of the first kind.
Let us have a look at inequality (\ref{inequality}). This is nothing but a condition
which guarantees that our spectrum is a pure real function. In addition
$$
\left|\frac{\lambda - \lambda_\zeta}{\zeta}\right| \le
\left|\frac{\lambda - \lambda_\zeta}{\zeta\cos\theta}\right| \le 1.
$$
It means that all functions in (\ref{isotr_K_part}) are real and we can omit
the part with imaginary unit $\I$.
After substitution
Eq. (\ref{isotr_K_part}) into 
Eq. (\ref{isotr_spectrum_init_1}) one can obtain
\begin{equation}
\label{isotr_spectrum_init}
F_{is}(\lambda)=\frac{4}{\pi} \frac{g}{v^3} B
\int\limits_{0}^{+\infty} \Phi(\zeta) \D\zeta
K\left(\sqrt{1-\left(\frac{\lambda - \lambda_\zeta}{\zeta} \right)^2} \right).
\end{equation}
Expressions (\ref{anisotr_spectrum_init}) and (\ref{isotr_spectrum_init}) are
valid for an arbitrary dispersion relation $\lambda_\zeta = \omega (k(\zeta)) v /g$.

\section{Deep water case}
For further study of the spectrum we have to
introduce the explicit form 
of the dispersion relation. Here we consider one of the most important cases,
i.\,e.~deep water surface gravity waves. In the case of a deep water $\omega_k = 
\sqrt{gk}$, correspondently $\lambda_\zeta = \sqrt{\zeta}$. Let us consider an isotropic case
\begin{equation}
\label{Integral}
F_{is}(\lambda)=\frac{4}{\pi} \frac{g}{v^3} B \int\limits_{0}^{+\infty}
\Phi(\zeta) \D\zeta
K\left(\sqrt{1-\left(\frac{\lambda - \sqrt{\zeta}}{\zeta} \right)^2} \right).
\end{equation}
Now we have to take into account condition (\ref{inequality})
\begin{equation}
\zeta^2  - \left(\lambda - \sqrt{\zeta} \right)^2 \ge 0.
\end{equation}
Solutions of this inequality give us integration domains
\begin{equation}
\label{int_limits}
\begin{array}{l}
\displaystyle
\rm{when} \;\; \lambda \le \frac{1}{4}\\
\displaystyle
\left(-\frac{1}{2}+ \sqrt{\frac{1}{4}+\lambda} 
\right)^2 < \zeta < \left( \frac{1}{2}- \sqrt{\frac{1}{4}-\lambda}\right)^2,\\
\displaystyle
\rm{when} \;\; \lambda > \frac{1}{4}\\
\displaystyle
\left(-\frac{1}{2}+ \sqrt{\frac{1}{4}+\lambda}
\right)^2 < \zeta < +\infty,
\end{array}
\end{equation}
so we have following expressions for spectra
\begin{eqnarray}
\mbox{when} \;\; \lambda \le \frac{1}{4},\nonumber\\
F_{is}(\lambda)=\frac{4}{\pi} \frac{g}{v^3} B \left[
\int\limits_{\left(-\frac{1}{2}+ \sqrt{\frac{1}{4}+\lambda} 
\right)^2}^{\left( \frac{1}{2}- \sqrt{\frac{1}{4}-\lambda}\right)^2} 
\Phi(\zeta)
 K \left( \sqrt{1-\left( \frac{\lambda - \sqrt{\zeta}}{\zeta} \right)^2}
\right) \D\zeta \right. +\label{isotr_first_domain}\\
  \left. +\int\limits_{\left(\frac{1}{2}+ \sqrt{\frac{1}{4}-\lambda}
\right)^2}
^{+\infty} \Phi(\zeta)K\left(\sqrt{1-\left(\frac{\lambda -
\sqrt{\zeta}}{\zeta} \right)^2} \right) \D\zeta\right];\nonumber
\end{eqnarray}
\begin{eqnarray}
\mbox{when} \;\; \lambda > \frac{1}{4},\nonumber\\
F_{is}(\lambda)=\frac{4}{\pi} \frac{g}{v^3} B\int\limits_{\left(-\frac{1}{2}+
\sqrt{\frac{1}{4}+\lambda} \right)^2}^{+\infty} \Phi(\zeta) \D\zeta
K\left(\sqrt{1-\left(\frac{\lambda - \sqrt{\zeta}}{\zeta} \right)^2} \right).\label{isotr_second_domain}
\end{eqnarray}
Let us explore special case, when $\Phi(\zeta)$ is a quickly going down to
zero function, e.g. Phillips spectrum $\Phi(\zeta)=1/\zeta^4$.
In this case we can take into account only the first term in
Eq.~(\ref{isotr_first_domain}) at $\lambda < 1/4$. Calculation
of the expansion of the spectrum correction is placed in Appendix~\ref{appendix_A} and (\ref{app:isotr})
gives us the following result
\begin{equation}
\label{iso_small_analytics}
F_{is}(\lambda) = F_{is 0} (1+\frac{1}{4}\lambda^2+...),
\end{equation}
where
\begin{equation}
F_{is 0}= 4 \pi \alpha \frac{v^5}{g^3} \frac{1}{\lambda^5},
\end{equation}
this is simply the Phillips spectrum Eq. (\ref{Phillips-isotropic}) at the
absence of Doppler effect for isotropic distribution.

For the one-directional spectra we can use exactly the same approach, but all
calculations become simpler.
The domains of integrations (which correspond to Eq.
(\ref{isotr_first_domain}) and 
Eq. (\ref{isotr_second_domain})) are the following
\begin{eqnarray}
\mbox{when} \;\; \lambda \le \frac{1}{4},\nonumber\\
F_{an}(\lambda)=\frac{2}{\pi} \frac{g}{v^3} B \left(
\int\limits_{\left(-\frac{1}{2}+ \sqrt{\frac{1}{4}+\lambda} 
\right)^2}^{\left( \frac{1}{2}- \sqrt{\frac{1}{4}-\lambda}\right)^2}
\frac{\Phi(\zeta)\D\zeta}{ \sqrt{1-\left( \frac{\lambda -
\sqrt{\zeta}}{\zeta} \right)^2}} + \right. \label{anisotr_first_domain}\\
\left. +\int\limits_{\left(\frac{1}{2}+ \sqrt{\frac{1}{4}-\lambda}
\right)^2}^{+\infty}
\frac{\Phi(\zeta) \D\zeta}{ \sqrt{1-\left(\frac{\lambda - \sqrt{\zeta}}{\zeta}
\right)^2}}\right);\nonumber
\end{eqnarray}
\begin{eqnarray}
\mbox{when} \;\; \lambda > \frac{1}{4},\nonumber\\
F_{an}(\lambda)=\frac{2}{\pi} \frac{g}{v^3} B \int\limits_{\left(-\frac{1}{2}+
\sqrt{\frac{1}{4}+\lambda} \right)^2}^{+\infty} 
\frac{\Phi(\zeta) \D\zeta}{\sqrt{1-\left(\frac{\lambda - \sqrt{\zeta}}{\zeta}
\right)^2}}.\label{anisotr_second_domain}
\end{eqnarray}

The detailed analysis of this case is situated in Appendix~\ref{appendix_A}
and (\ref{app:anisotr}) gives us the following formula
\begin{equation}
\label{anis_small_analytics}
F_{an}(\lambda) = F_{an 0} (1+\frac{1}{2}\lambda^2+...),
\end{equation}
where
\begin{equation}
F_{an 0}= 4 \alpha \frac{v^5}{g^3} \frac{1}{\lambda^5},
\end{equation}
this is just the Phillips spectrum Eq. (\ref{Phillips-anisotropic}) at the
absence of Doppler effect for strongly anisotropic distribution.

Let us try to evaluate the range of applicability of the obtained results. As it was
postulated in Eq. (\ref{lambda_sence}) we can state
$$
\lambda = \frac{\omega}{\omega_p} \frac{v}{c_p} < \frac{1}{4}.
$$
Thus, for example, if we take $\varepsilon = v/c_p = 0.05$ (a reasonable value of wind drift current \cite{Banner1990}, or if
we consider the influence of high frequency wave on the far tail of spectrum) one
can obtain the following inequality
\begin{equation}
\label{applicability_condition}
\omega < 5 \omega_p.
\end{equation}

If we consider a Doppler shift due to long waves corresponding to the vicinity of spectral maximum the result will be different. Following (\ref{Doppler_velocity_peak}) $v = A_p \omega_p$.

In this case the physical meaning of our dimensionless variables is quite straightforward
\begin{equation}
\zeta = \mu^2 \frac{k}{k_p},\;\; \lambda = \mu \frac{\omega}{\omega_p},
\end{equation}
here $k_p$ and $\omega_p$ are wavenumber and circular frequency corresponding to spectrum peak,
and $\mu = A_p k_p$, characteristic steepness of spectrum peak.
Usually this value is about $\mu \simeq 0.1$. This gives us estimations
\begin{equation}
\omega < 2.5 \omega_p.
\end{equation}

Although this more or less corresponds to the frequency bandwidth of many experiments, we need
to find an asymptotic of Phillips frequency spectrum $F(\lambda)$ at high frequencies
$$
\lambda \gg 1/4.
$$
This condition simplifies situation significantly. In the isotropic case one can get
\begin{equation}
\label{isotr_far}
F_{is}(\lambda)\simeq\frac{4}{\pi} \frac{g}{v^3} B \int\limits_{\lambda}^{+\infty} \Phi(\zeta) \D\zeta
K\left(\sqrt{1-\left(\frac{\lambda}{\zeta} \right)^2} \right).
\end{equation}
In the anisotropic case
\begin{equation}
\label{anisotr_far}
F_{an}(\lambda)\simeq\frac{2}{\pi} \frac{g}{v^3} B \int\limits_{\lambda}^{+\infty} 
\frac{\Phi(\zeta) \D\zeta}{\sqrt{1-\left(\frac{\lambda}{\zeta}\right)^2}}.
\end{equation}
Here we left only term of the order of $\lambda$ under square root sign having supposed a fast decay of spectrum $\Phi(\zeta)$.
An obvious substitution $x = \lambda/\zeta$ in the case of Phillips spectrum $\Phi(\zeta) = \zeta^{-4}$
immediately leads to answers
\begin{equation}
\label{isotr_far_result}
F_{is}(\lambda)\simeq\frac{4}{\pi} \frac{\alpha v^5}{g^3} \frac{1}{\lambda^3}\int\limits_{0}^{1} x^2 \D x
K\left(\sqrt{1-x^2} \right) = \frac{\alpha v^5}{g^3}\frac{\pi}{4}\frac{1}{\lambda^3},
\end{equation}
\begin{equation}
\label{anisotr_far_result}
F_{an}(\lambda)\simeq\frac{2}{\pi} \frac{\alpha v^5}{g^3} \frac{1}{\lambda^3}\int\limits_{0}^{1} \frac{x^2 \D x}
{\sqrt{1-x^2}} = \frac{\alpha v^5}{g^3}\frac{1}{2}\frac{1}{\lambda^3},
\end{equation}
for isotropic and anisotropic spectra respectively.

As we can see slopes of the spectra changed and one could expect to find in the middle range of frequencies
long enough domain of $F(\omega) \sim \omega^{-4} \sim \lambda^{-4}$ spectrum dependence.
To develop this hypothesis we need to perform numerical calculation of integrals
(\ref{isotr_first_domain}), (\ref{isotr_second_domain}) and (\ref{anisotr_first_domain}), (\ref{anisotr_second_domain})
in the whole range of parameters.

\section{Numerical results}
At the first stage let us compare results in the case $\lambda < 1/4$ with our estimations.
\begin{figure}[hbt]
\includegraphics[width=12.0cm]{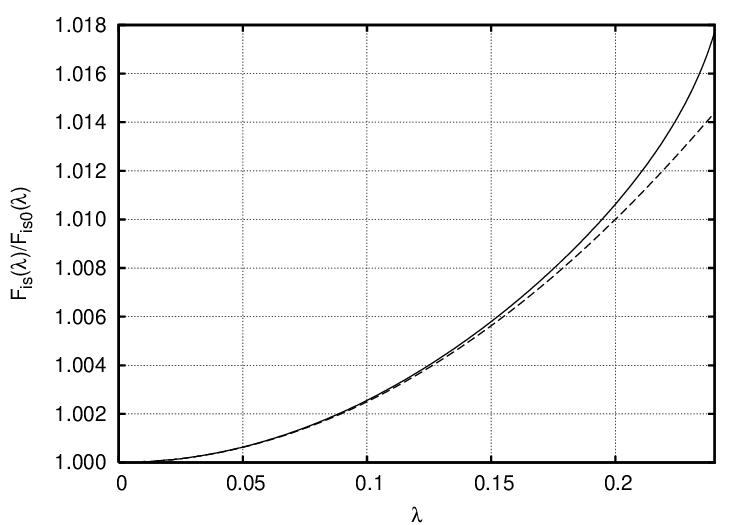}
\caption{\label{Iso_comp} Calculated spectrum in isotropic case
compensated with corresponding Phillips spectrum $F_{is}(\lambda)/F_{is 0}(\lambda)$ (solid line)
and analytic estimation (\ref{iso_small_analytics}) in the case of $\lambda < 1/4$ (dashed line).}
\end{figure}
\begin{figure}[hbt]
\includegraphics[width=12.0cm]{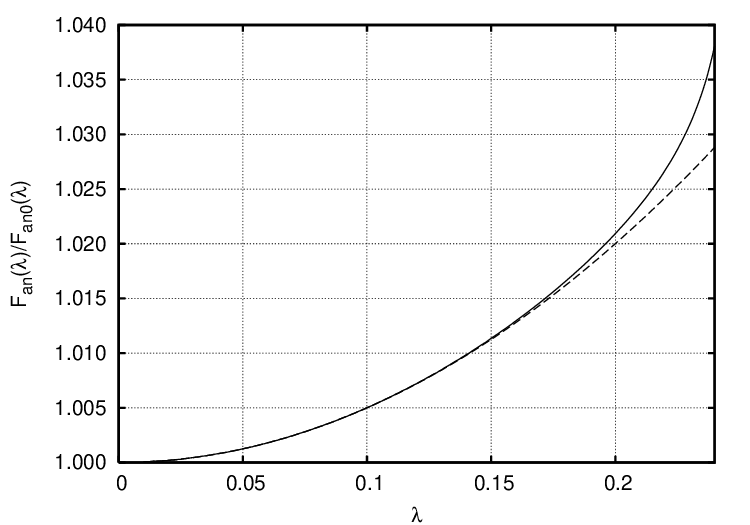}
\caption{\label{Aniso_comp} Calculated spectrum in strongly anisotropic case
compensated with corresponding Phillips spectrum $F_{an}(\lambda)/F_{an 0}(\lambda)$ (solid line)
and analytic estimation (\ref{anis_small_analytics}) in the case of $\lambda < 1/4$ (dashed line).}
\end{figure}
Results of these calculations are represented in Fig.~\ref{Iso_comp} and Fig.~\ref{Aniso_comp} for isotropic
and anisotropic cases. One can see, that numerical calculation gives us the same result --- there is a very small deviation
from Phillips spectra. Discrepancies of analytic and numerical results at the edge of the domain ($\lambda \simeq 1/4$)
shows that we cannot use only several terms expansions and have to take into account further terms in square root
decompositions used in Appendix~\ref{appendix_A}.

To understand results at the whole range of parameters it is convenient to use double logarithm representation.
\begin{figure}[hbt]
\includegraphics[width=12.0cm]{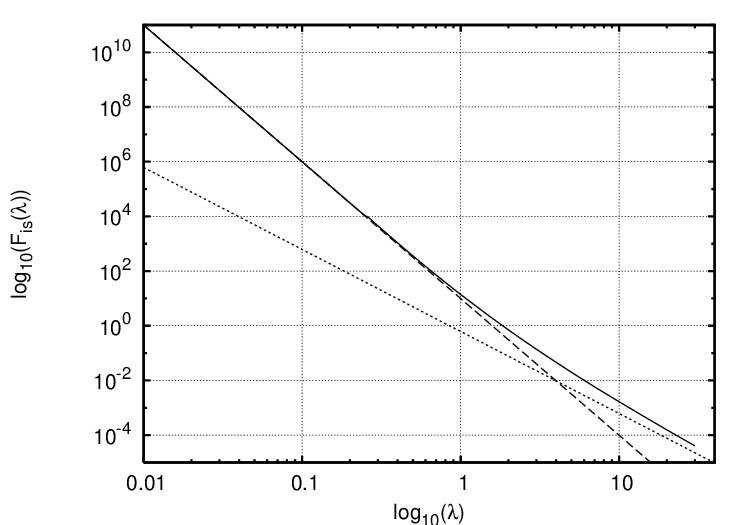}
\caption{\label{Iso_log_all} Calculated spectrum in isotropic case
$F_{is}(\lambda)$ (solid line), Phillips spectrum $F_{is 0}(\lambda)$ (dashed line)
and asymptotic (\ref{isotr_far_result}) corresponding to $\lambda \gg 1/4$ (dotted line).}
\end{figure}
\begin{figure}[hbt]
\includegraphics[width=12.0cm]{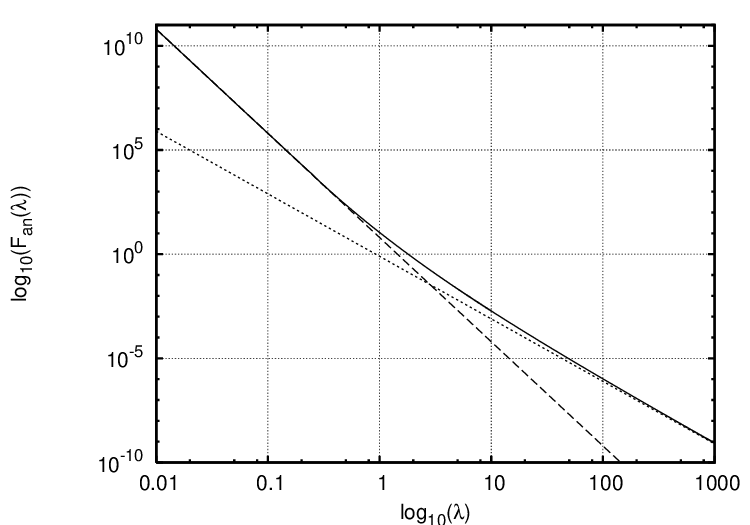}
\caption{\label{Aniso_log_all} The calculated spectrum in the strongly anisotropic case
$F_{an}(\lambda)$ (solid line), the Phillips spectrum $F_{an 0}(\lambda)$ (dashed line)
and the asymptotic (\ref{anisotr_far_result}) corresponding to $\lambda \gg 1/4$ (dotted line).}
\end{figure}
One can see in Fig.~\ref{Iso_log_all} and Fig.~\ref{Aniso_log_all} that Phillips spectrum remains untouched
in a wide range of parameters and then transforms to asymptotic (\ref{isotr_far_result}) and
(\ref{anisotr_far_result}) respectively.

Now we can consider the interesting problem of Doppler distortion due to the influence of spectrum peak.
In this special case we are interested in the region of parameters $ 1 < \lambda < 10$ corresponding
to the switch of the spectrum slope. Following our results in such a situation, one could expect
formation of $\omega^{-4} \sim \lambda^{-4}$ spectrum due to transfer from Phillips $\lambda^{-5}$
to $\lambda^{-3}$ tail. Let us consider our results in details.
\begin{figure}[hbt]
\includegraphics[width=12.0cm]{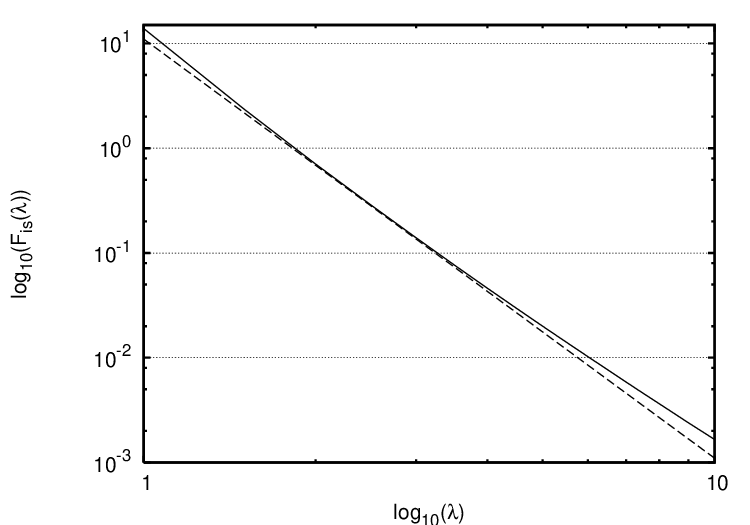}
\caption{\label{Iso_log_small} Calculated spectrum in isotropic case
$F_{is}(\lambda)$ (solid line) and frequently observed spectrum $\lambda^{-4}$ (dashed line).
The spectrum slope is close to $\omega^{-4}$ for $2 < \lambda < 3$.}
\end{figure}
\begin{figure}[hbt]
\includegraphics[width=12.0cm]{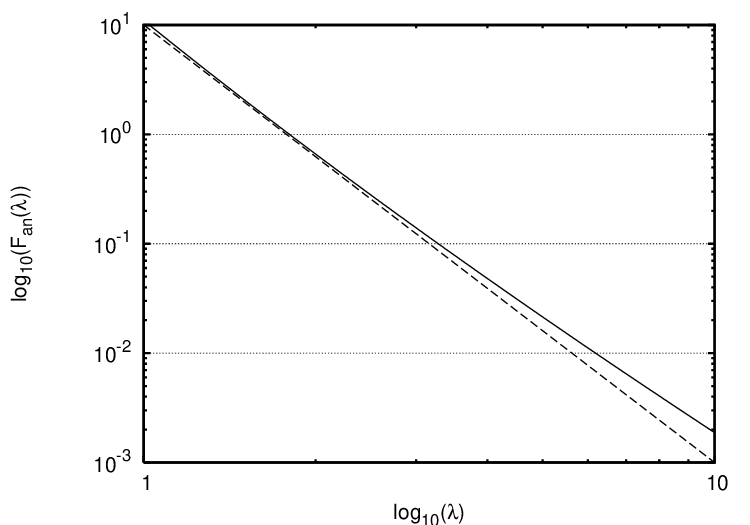}
\caption{\label{Aniso_log_small} Calculated spectrum in strongly anisotropic case
$F_{an}(\lambda)$ (solid line) and frequently observed spectrum $\lambda^{-4}$ (dashed line).
The spectrum slope is close to $\omega^{-4}$ for $1.5 < \lambda < 2$.}
\end{figure}
It is clear from Fig.~\ref{Iso_log_small} and Fig.~\ref{Aniso_log_small} that
there is no such a region. Universal spectra $\omega^{-4}$ only touch our curve. It means
that Doppler distortion due to spectral peak waves influence cannot be considered
as a mechanism of transformation of the analytic Phillips spectra $\omega^{-5}$ to the frequently
observed $\omega^{-4}$.

From the Fig.~\ref{Iso_log_all} and \ref{Aniso_log_all} one can notice that we have
$\omega^{-5}$ tail at least up to $\lambda = 0.5$. Following physical meaning of $\lambda$
(\ref{lambda_sence}) we have that $\omega^{-5}$ tail stays the same at least for
$$
\omega \le 5\omega_p,
$$
in the case when spectrum peak steepness is about $\mu \simeq 0.1$. In fact we observe tail close to $\omega^{-4}$
only in the vicinity of the points $\lambda\simeq 2$, which corresponds at such value of steepness to
$\omega = 20 \omega_p$ --- beyond the frequency bandwidth of even current state of the art water tank experiments~\cite{Lukaschuk2007}.

We would like to stress, that change of
spectrum peak steepness does not affect general results of the paper, but shifts the frequency region where
we observe turn from $\omega^{-5}$ to $\omega^{-3}$ tail.

In spite of the simplicity of formulae under consideration they are not suitable for reliable
numerical calculations
(especially in the case $\lambda >> 1/4$) using common numerical calculation packets, because of points
on both sides of integration interval and inside of it which give singularities of expression under
integral sign, causing very slow convergence of algorithm.
Nevertheless all these obstacles can be avoided or eliminated using standard numerical analysis
techniques. As an examples we represented direct formula for integration in the $\lambda > 1/4$ case
of strongly anisotropic spectrum in Appendix~\ref{appendix_B}

\section{Conclusion}
In this paper we considered the influence of the Doppler distortion on the surface waves spectral tails.
The general formula~(\ref{spectrum_dim}) for an arbitrary spatial spectra and arbitrary dispersion relation was derived. The case of dispersion relation corresponding to surface gravity waves was considered in detail.
The formulae for isotropic~(\ref{isotr_first_domain}-\ref{isotr_second_domain})
and the strongly anisotropic~(\ref{anisotr_first_domain}-\ref{anisotr_second_domain}) spectra were obtained.
For the special case of Phillips spectra $\sim k^{-4}$ we got two asymptotics:
$\omega^{-5}$ in the frequency region close ($\omega < 2.5\omega_p$ in the case of spetrum peak influence)
to the spectral maximum~(\ref{iso_small_analytics}),(\ref{anis_small_analytics}) and
$\omega^{-3}$ for high frequencies~(\ref{anisotr_far}), (\ref{isotr_far}).

We numerically calculated the spectra in the wide range of parameters and gave an explanation
of the results considering the range of applicability of assumptions used.

\appendix

\section{Calculation of the spectra relations and Doppler velocity due to spectral peak's influence.}
\label{appendix_0}
Following \cite{Zakharov1992} or \cite{DKZ2003} we reformulate our problem using the Hamiltonian
formalism.

Let us consider the potential flow of an ideal incompressible fluid
of infinite depth and with a free surface.
We use standard notations for velocity potential $\phi(\vec r, z, t),\vec r = (x,y); v= \nabla\phi$ and surface elevation
$\xi(\vec r, t)$. Fluid flow is irrotational $\Delta\phi = 0$. The total energy of the system can be
represented in the following form
$$
H = T + U,
$$
\begin{equation}
T = \frac{1}{2} \int \D^2 r \int \limits_{-\infty}^{\xi} (\nabla \phi)^2 \D z,
\end{equation}
\begin{equation}
U = \frac{1}{2} g \int \xi^2 \D^2 r,
\end{equation}
where $g$ -- is the gravity acceleration. It was shown \cite{Zakharov-68} that under these assumptions
the fluid is a Hamiltonian system
\begin{equation}
\label{Hamiltonian_equations}
\frac{\partial \xi}{\partial t} = \frac{\delta H}{\delta \psi}, \;\;\;\;
\frac{\partial \psi}{\partial t} = - \frac{\delta H}{\delta \xi},
\end{equation}
where $\psi = \phi (\vec r, \xi (\vec r,t), t)$ is a velocity potential on the surface of the fluid. In order to
calculate the value of $\psi$ we have to solve the Laplas equation in the domain with varying
surface $\eta$. This is a difficult problem. One can simplify the situation, using the expansion
of the Hamiltonian in powers of ''steepness''. If we limit ourselves only to the linear part
\begin{equation}
\label{Hamiltonian}
H = \frac{1}{2}\int\left( g \xi^2 + \psi \hat k  \psi \right) \D^2 r.
\end{equation}
Here $\hat k f(\vec r) = \hat F^{-1} [|k|\hat F[f(\vec r)]]$ --- is the linear operator corresponding to
multiplying of Fourier harmonics by absolute value of the wavenumber $\vec k$.
In this case linear dynamical equations (\ref{Hamiltonian_equations}) acquire the following form
\begin{equation}
\label{eta_psi_system}
\dot \xi = \hat k  \psi,\;\;\;
\dot \psi = - g\xi.
\end{equation}
Let us consider velocity field in $(x,y)$ plane which results from the presence of spectral peak wave.
The monochromatic spectral peak wave can be represented as follows
$$
\xi_p (\vec r, t) = A_p \cos ((\vec k_p \vec r) - \omega_p t).
$$
Taking into account one of Hamiltonian equations (\ref{eta_psi_system}),
velocity field in $\vec r = (x,y)$ plane is given by the formula
$$
\vec v_p (\vec r, t) = \nabla \psi (\vec r, t) =
- g \int\limits_{0}^{t} \nabla \xi_p (\vec r, t')\D t'.
$$
Finally, Doppler velocity in the direction of $\vec k_p$ is the following
\begin{equation}
\label{Doppler_velocity_peak}
v_p (t) = \frac{g k_p A_p}{\omega_p} \cos ((\vec k_p \vec r) - \omega_p t) =
A_p \omega_p \cos ((\vec k_p \vec r) - \omega_p t).
\end{equation}

Fourier harmonics of the real functions $\psi$ and $\xi$ have Hermitian symmetry, so
it is convenient to introduce the canonical variables $a_{\vec k}$ as shown below
\begin{equation}
\label{a_k_substitution}
a_{\vec k} = \sqrt \frac{\omega_k}{2k} \eta_{\vec k} + \I \sqrt \frac{k}{2\omega_k} \psi_{\vec k},
\end{equation}
where
\begin{equation}
\label{dispersion_relation}
\omega_k = \sqrt {g k}.
\end{equation}
This is the dispersion relation for the case of gravity waves on the surface of the fluid of infinite depth.
Similar formulas can be derived in the case of finite depth \cite{Zakharov-1999}. We should stress,
that expression (\ref{a_k_substitution}) is valid for arbitrary dispersion relation.

With these variables the equations (\ref{Hamiltonian_equations}) take the following form
\begin{equation}
\label{Hamiltonian_eqs_canonical}
\dot a_{\vec k} = -i \frac{\delta H}{\delta a_{\vec k}^{*}}.
\end{equation}
The most important feature of these variables is the absence of Hermitian symmetry: it means that if
there is only a monochromatic wave with wave vector $\vec k$, then only $a_{\vec k}$ harmonics
will have nonzero value.

One can introduce spectra
\begin{equation}
\langle a_{\vec k} a_{\vec k'}^{*}\rangle = N_{\vec k} \delta (\vec k - \vec k'),\;\;\;
\langle a_{\vec k, \omega} a_{\vec k', \omega'}^{*}\rangle = N_{\vec k, \omega} \delta (\vec k - \vec k')
\delta (\omega - \omega').
\end{equation}
For the spectrum related with $N_{\vec k}$ Hasselmann equation (\ref{Hasselmann}) is written.

Using representation of $\xi_{\vec k}$ in terms of $a_{\vec k}$
$$
\xi_{k} = \sqrt{\frac{k}{2\omega_k}}\left(a_{\vec k} + a_{-\vec k}^*\right),
$$
and the definition of the spatial spectrum of the waves, one can get
\begin{eqnarray}
\Phi(\vec k)\delta(\vec k - \vec k') = \langle\xi_{\vec k}\xi_{\vec k}^{*}\rangle =
\frac{k}{2\omega}\left(N_{\vec k} + N_{-\vec k}\right)\delta(\vec k - \vec k'),\\
\Phi(\vec k) = \frac{k}{2\omega}\left(N_{\vec k} + N_{-\vec k}\right).
\end{eqnarray}
In the same way for spatial-temporal spectrum (\ref{spectrum})
\begin{equation}
\label{Chi_N_kw}
X(\vec k, \omega) = \frac{k}{2\omega}\left(N_{\vec k, \omega} + N_{-\vec k, -\omega}\right).
\end{equation}
In the linear approximation
\begin{equation}
N_{\vec k, \omega} = N_{\vec k} \delta(\omega - \omega_k),
\end{equation}
and relation (\ref{Chi_N_kw}) can be written as
\begin{equation}
\label{Chi_N_k}
X(\vec k, \omega) = \frac{k}{2\omega}\left(N_{\vec k}\delta(\omega - \omega_k) +
N_{-\vec k}\delta(\omega + \omega_k)\right).
\end{equation}
Let's consider two limiting cases of the wave field:
\begin{itemize}
\item
Spectrum is symmetric on $\vec k$;
\item
Spectrum is unidirectional on $\vec k$.
\end{itemize}

For isotropic spectrum
$$
N_{\vec k} = N_{-\vec k},\;\;\; \Phi(k) = \frac{k}{\omega_k}N_{k},\;\;\;
X_{is}(k, \omega) = \frac{k}{2\omega}N_{k}\left(\delta(\omega - \omega_k) +
\delta(\omega + \omega_k)\right);
$$
\begin{equation}
\label{Chi_Phi_iso}
X_{is}(k, \omega) = \frac{1}{2}\Phi_{k}\left(\delta(\omega - \omega_k) +
\delta(\omega + \omega_k)\right).
\end{equation}

For unidirectional spectrum (with direction corresponding to angle $\theta_0$)
$$
N_{\vec k} = N_{k}\delta(\theta-\theta_0),\; N_{-\vec k} = N_{k}\delta(\theta-\pi-\theta_0),
$$
$$
X_{an}(\vec k, \omega) = \frac{k}{2\omega}N_{k}\left(\delta(\theta-\theta_0)\delta(\omega - \omega_k) +
\delta(\theta-\pi-\theta_0)\delta(\omega + \omega_k)\right).
$$
We consider positive-frequency part:
$$
\Phi(\vec k) = \frac{k}{2\omega_k}N_{\vec k},\;\;\;
X_{an}(\vec k, \omega) = \frac{k}{2\omega}N_{\vec k}\delta(\omega - \omega_k);
$$
\begin{equation}
\label{Chi_Phi_aniso}
X_{an}(\vec k, \omega) = \Phi(\vec k)\delta(\omega - \omega_k).
\end{equation}

\section{Calculations of the spectra decompositions. $\lambda < 1/4$.}
\label{appendix_A}
As the first step, we consider slightly more complex case of isotropic spectrum.
Let us perform the following substitution
\begin{equation}
x=\frac{\lambda - \sqrt{\zeta}}{\zeta}.
\end{equation}
In this case in our domain of integration ($-1< x < +1$) we can obtain the
inverse representation
\begin{equation}
\zeta=\frac{1}{2 x^2}\left[ 1 - \sqrt{ 1+4 x \lambda} + 2 x \lambda
\right],
\end{equation}
and further
\begin{equation}
\zeta(x) \approx \frac{1}{2 x^2} \left[ 2 \lambda^2 x^2 - 4 \lambda^3 x^3 +10
\lambda^4 x^4 \right]
=\lambda^2 - 2 \lambda^3 x + 5 \lambda^4 x^2.
\end{equation}
For substitution $d \zeta = \zeta' (x) dx$ we have to calculate the first
derivative
\begin{eqnarray}
\zeta'(x) = -\frac{1}{x^3} - \frac{\lambda}{x^2 \sqrt{1+4 \lambda x}} +
\frac{\sqrt{1+4 \lambda x}}{x^3} -\nonumber\\
-\frac{\lambda}{x^2} \approx
-\frac{1}{x^3} - \frac{\lambda}{x^2} \left[ 1- 2 \lambda x + 6 \lambda^2 x^2
-20 \lambda^3 x^3\right. +\\ 
+\left.70 \lambda^4 x^4 \right]
+ \frac{1}{x^3} \left[ 1+ 2\lambda x - 2 \lambda^2 x^2 + 4 \lambda^3 x^3
-\right.\nonumber\\
-\left. 10 \lambda^4 x^4 + 
28 \lambda^5 x^5 \right] -\frac{\lambda}{x^2}=\nonumber\\
= - 2 \lambda^3 \left( 1-5 \lambda x + 21 \lambda^2 x^2 \right).\nonumber
\end{eqnarray}
\begin{eqnarray}
\Phi(\zeta(x)) = \Phi (\lambda^2 (1- 2 \lambda x + 5 \lambda^2 x^2)) \approx
\nonumber\\
\approx \Phi(\lambda^2) - \frac{\Phi'(\lambda^2)}{1!} 2 \lambda^3 x +\\
+ \frac{\Phi'(\lambda^2)}{1!} 5\lambda^4 x^2  
+\frac{\Phi''(\lambda^2)}{2!} 4 \lambda^6 x^2.\nonumber
\end{eqnarray}
Now we can calculate the integral (\ref{isotr_first_domain}) with Phillips
spectrum ($\Phi(\zeta)=1/\zeta^4$)
\begin{equation}
\label{difference}
F_{is}(\lambda)= \frac{4}{\pi} \frac{g}{v^3} B  
\int\limits_{-1}^{+1} 
\Phi(\zeta(x)) \zeta'(x) K (\sqrt{1-x^2}) \D x,
\end{equation}
Taking into account several well known results \cite{GradshteynRyzhik}
$$
\int\limits_{- 1}^{+ 1}K \left(\sqrt{1-x^2} \right)\D x=\frac{\pi^2}{2},
$$
$$
\int\limits_{- 1}^{+ 1}xK \left(\sqrt{1-x^2} \right)\D x \equiv 0,
$$
$$
\int\limits_{-1}^{+ 1}x^2 K \left(\sqrt{1-x^2} \right)\D x=\int\limits_{0}^{+1}E \left(\sqrt{1-x^2}
\right)\D x=\frac{\pi^2}{8},
$$
where $E$ is the complete elliptic integral of the second kind, one can obtain the final result
\begin{equation}
\label{app:isotr}
F_{is}(\lambda) = 4 \pi \alpha \frac{v^5}{g^3} \frac{1}{\lambda^5}
(1+\frac{1}{4}\lambda^2+...).
\end{equation}

Calculations of the spectrum decomposition for strongly anisotropic
case up to expression Eq. (\ref{difference}) are just the same.
Integral (\ref{anisotr_first_domain}) takes the form
\begin{equation}
F_{an}(\lambda)= \frac{2}{\pi} \frac{g}{v^3} B \int\limits_{-1}^{+1} 
\frac{\Phi(\zeta(x)) \zeta'(x)}{ \sqrt{1-x^2}} \D x,
\end{equation}
The final result is
\begin{equation}
\label{app:anisotr}
F_{an}(\lambda) = 2 \alpha \frac{v^5}{g^3} \frac{1}{\lambda^5}
(1+\frac{1}{2}\lambda^2+...).
\end{equation}

\section{Numerical calculation of the strongly anisotropic spectra at $\lambda > 1/4$.}
\label{appendix_B}
Here we consider Phillips spatial spectrum $\zeta^{-4}$ but all results are independent with respect to
the power of spatial spectrum. We start from (\ref{anisotr_second_domain}). Here and further we omit all constants before
integral sign
\begin{equation}
F(\lambda) = \int\limits_{\left(\sqrt{\frac{1}{4} + \lambda} - \frac{1}{2}\right)^2}^{+\infty}
\frac{\zeta^{-4} \D\zeta}{\sqrt{1 - \left(\frac{\lambda - \sqrt{\zeta}}{\zeta}\right)^2}}.
\end{equation}
After substitution $\mu = 1/\sqrt{\zeta}$ one gets
\begin{equation}
F(\lambda) = 2\int\limits_{0}^{\left(\sqrt{\frac{1}{4} + \lambda} - \frac{1}{2}\right)^{-1}}
\frac{\mu^{5} \D\mu}{\sqrt{1 - \left(\lambda\mu^2 - \mu\right)^2}}.
\end{equation}
The upper limit of integration gives us a singularity. Usual technique for calculation of this integral consists
in excluding of this special point in some way (see for example \cite{Fedorenko1994}.
In our case one can do this in a very simple and effective way. Our integral
can be represented as follows
\begin{equation}
F(\lambda) = 2\int\limits_{0}^{\mu_2}
\frac{\mu^{5} \D\mu}
{\sqrt{\lambda\left(\mu_1 -\mu\right)\left(\mu - \mu_2\right)\left(1+\lambda\mu^2 -\mu\right)}},
\end{equation}
where we introduced two roots of polynomial under square root sign
\begin{eqnarray*}
\mu_1 &=& \frac{1}{\lambda}\left(\frac{1}{2} - \sqrt{\frac{1}{4} + \lambda}\right),\\
\mu_2 &=& \frac{1}{\lambda}\left(\frac{1}{2} + \sqrt{\frac{1}{4} + \lambda}\right).
\end{eqnarray*}
Using substitution $\nu = \sqrt{\mu_2 - \mu}$ one can get
\begin{equation}
F(\lambda) = \frac{4}{\sqrt{\lambda}}\int\limits_{0}^{\sqrt{\mu_2}}
\frac{(\mu_2 - \nu^2)^{5} \D\nu}
{\sqrt{\left(\mu_1 - \mu_2 +\nu^2\right)\left(1+\lambda(\mu_2 - \nu^2)^2 -\mu_2 + \nu^2\right)}}.
\end{equation}
This integral can be calculated by any standard tool of numerical mathematics, for example Simpson rule
\cite{Fedorenko1994}, \cite{NumericalRecipes}. In this
case even 32 points in domain give us relative error less $10^{-6}$ with respect to 16 points.
If we would try to calculate our spectrum from the very first formula, to achieve such an accuracy we
have to use more than $10^{6}$ points in integration domain (because even Simpson rule, usually the method of the 
third order of accuracy, has only the first order in the vicinity of singular point). With modern computers and such a simple
functions as a square root under integration sign it is not very fast but bearable calculation. But if we have to
calculate integral of special function (for instance, the complete elliptic integral of the first kind)
it will take an enormous amount of time.

%\begin{acknowledgements}
\section*{Acknowledgments}
The author gratefully wishes to acknowledge the following contributions:
formulation of this problem and many enlightening
and helpful discussions by Prof. V.\,E.~Zakharov; useful discussions about numerical methods by
Prof. M.\,G.~Stepanov, Department of Mathematics, The University of Arizona;
Landau Scholarship Committee; grant
RFBR 06-01-00665-a; the Programme 
``Fundamental problems of Nonlinear dynamics'' from the RAS Presidium and ``Leading 
Scientific Schools of Russia'' grant NSh-7550.2006.2.

Also the author wishes to thank creators of GNU Scientific Library \cite{GSL} for
this useful and free software.
%\end{acknowledgements}


\begin{thebibliography}{99}
\bibitem{Phillips1958}
O.\,M.~Phillips,
%The equilibrium range in the spectrum of wind generated waves.
J. Fluid Mech. 4 (1958) 426.

\bibitem{Toba1973}
Y.~Toba,
%Local balance in the air-sea boundary processes Part II: 
%On the spectrum of wind generated waves.
J. Oceanogr. Soc. Japan  29 (1973) 209.

\bibitem{Donelan1985}
M.\,A.~Donelan, J.~Hamilton, and W.\,H.~Hui,
%Directional spectra of wind generated waves.
Phil. Trans. R. Soc. London A315 (1985) 509.

\bibitem{Hwang1999}
P.\,A.~Hwang at al., J. Phys. Oceanogr. 30 (1999) 2753.

\bibitem{Banner1990}
M.\,L.~Banner,
%Equilibrium Spectra of Wind Generated Waves.
Journ. of Physical Oceanography 20 (1990) 966.

\bibitem{Hasselmann1962}
K.~Hasselmann,
%On the nonlinear energy transfer in a gravity wave spectrum. Part 1. 
J. Fluid Mech. 12 (1962) 481.

\bibitem{Hasselmann1963}
K.~Hasselmann,
%On the nonlinear energy transfer in a gravity wave spectrum. Part 2 and Part 3. 
J. Fluid Mech. 15 (1963) 273 and 385.

\bibitem{Zakharov1966}
V.\,E.~Zakharov  and N.\,N.~Filonenko, J. Appl. Mech. Tech. Phys 4 (1966)

\bibitem{Zakharov1967}
V.\,E.~Zakharov  and N.\,N.~Filonenko, Dokl. Acad. Nauk SSSR 170 (1967) 881.\\
(V.\,E.~Zakharov and N.\,N.~Filonenko,
%Energy spectrum for stochastic oscillations of the surface of a liquid.
Soviet Phys. Dokl., 11 (1968) 881.)

\bibitem{Craik1987}
A.\,D.\,D.~Craik,
%Interaction of a short-wave field with a dominant long wave in deep water: derivation from Zakharov's spectral formulation.
Journ. Austral. Math. Soc. Ser. B 29 (1987) 430.

\bibitem{Zakharov1992}
V\,E.~Zakharov, G.~Falkovich, V.~Lvov, Kolmogorov spectra of turbulence I (Springer, Berlin), 1992\\
ISBN: 3-540-54533-6.

\bibitem{PZ1996}
A.\,N.~Pushkarev and V.\,E.~Zakharov, Phys. Rev. Lett. 76 (1996).

\bibitem{Onorato2002}
M.~Onorato, A.\,R.~Osborne, M.~Serio at al., Phys.\,Rev.\,Lett. 89 (2002) 144501.

\bibitem{DKZ2003}
A.\,I.~Dyachenko, A.\,O.~Korotkevich, V.\,E.~Zakharov,
%Weak turbulence of gravity waves,
JETP Lett. 77 (2003) 546.

\bibitem{DKZ2004}
A.\,I.~Dyachenko, A.\,O.~Korotkevich, V.\,E.~Zakharov,
%Weak turbulent Kolmogorov spectrum for surface gravity waves. 
Phys. Rev. Lett. 92 (2004) 134501.

\bibitem{ZKPD2005}
V.\,E.~Zakharov, A.\,O.~Korotkevich, A.\,N.~Pushkarev, and A.\,I.~Dyachenko,
% Mesoscopic Wave Turbulence,
JETP Lett. 82 (2005) 487.

\bibitem{KPRZ2007}
A.\,O.~Korotkevich, A.\,N.~Pushkarev, D.~Resio, and V.\,E.~Zakharov,
%Numerical Verification of the Weak Turbulent Model for Swell Evolution,
Eur. J. Mech. B/Fluids in press (2007).

\bibitem{ZKPR2007}
V.\,E.~Zakharov, A.\,O.~Korotkevich, A.\,N.~Pushkarev, and D.~Resio,
%Coexistence of weak and strong wave turbulence in a swell propagation,
Phys. Rev. Lett. 99 (2007) 164501.

\bibitem{Phillips1986}
O.\,M.~Phillips,
%Spectral and statistical properties of the equilibrium range in wind-generated gravity waves,
J. Fluid Mech. 156 (1986) 505.

\bibitem{Donelan1998}
M.\,A.~Donelan, %Air-water exchange processes.
In: Physical Processes in Oceans
and Lakes 
(J. Imberger, Ed.), AGU Coastal and Estuarine Studies, 54 (1998) 19.

\bibitem{KKZ1975}
S.\,A.~Kitaigorodskii, V.\,P.~Krasitskii, M.\,M.~Zaslavskii,
%On Phillips' theory of equilibrium range in the spectra of wind-generated gravity waves. 
J. Phys. Oceanogr. 5 (1975) 410.

\bibitem{Phillips1981}
O.\,M.~Phillips,
%The dispersion of short wavelets in the presence of a dominant long wave.
J. Fluid Mech. 107 (1981) 465.

\bibitem{Kuznetsov2004}
E.\,A.~Kuznetsov,
%On Phillips' theory of equilibrium range in the spectra of wind-generated gravity waves. 
JETP Letters 80 (2004) 83.

\bibitem{GradshteynRyzhik}
I.\,S.~Gradshteyn, I.\,M.~Ryzhik, A.~Jeffrey, and D.~Zwillinger,
Table of Integrals, Series, and Products, Sixth Edition, Academic Press, 2000\\
ISBN: 0-12-294757-6.

\bibitem{Lukaschuk2007}
P.~Denissenko, S.~Lukaschuk, and S.~Nazarenko, Phys. Rev. Lett. 99 (2007) 014501.

\bibitem{Zakharov-68}
V.\,E. Zakharov, J. Appl. Mech. Tech. Phys. 2 (1968) 190.

\bibitem{Zakharov-1999}
V.\,E. Zakharov, Eur. J. Mech. B 18 (1999) 327.

\bibitem{Fedorenko1994}
R.\,P.~Fedorenko, Introduction to the computational physics, Moscow Institute of Physics and Technology Press, 1994\\
ISBN: 5-7417-0002-0.

\bibitem{NumericalRecipes}
W.\,H.~Press, B.\,P.~Flannery, S.\,A.~Teukolsky, and W.\,T.~Vetterling,
"Numerical Recipes: The Art of Scientific Computing", Cambridge University Press, http://nr.com\\
ISBN: 0-5214-3108-5.

\bibitem{GSL}
GNU Scientific Library, http://www.gnu.org/software/gsl/

\end{thebibliography}
\end{document}